\documentclass[aps,prl,twocolumn,superscriptaddress,showpacs,,longbibliography]{revtex4-2}
\usepackage{amsmath,amssymb,amsfonts,bbm,float,graphics,epsfig,epstopdf,color,verbatim,tabularx,bm,multirow,hyperref,ulem,times,subfigure}
\hypersetup{
  colorlinks, linkcolor={blue},
  citecolor={blue}, urlcolor={blue}
}

\DeclareMathOperator{\Tr}{Tr}

\begin{document}
\title{High-efficiency quantum Monte Carlo algorithm for extracting entanglement entropy in interacting fermion systems}

\author{Weilun Jiang}
\affiliation{State Key Laboratory of Quantum Optics Technologies and Devices, Institute of Opto-Electronics, Shanxi University, Taiyuan, 030006, China }
\affiliation{Collaborative Innovation Center of Extreme Optics, Shanxi University, Taiyuan, 030006, China }

\author{Gaopei Pan}
\email{gaopei.pan@uni-wuerzburg.de}
\affiliation{Institut für Theoretische Physik und Astrophysik and Würzburg-Dresden Cluster of Excellence ct.qmat, Universität Würzburg, 97074 Würzburg, Germany}

\author{Zhe Wang}
\affiliation{Department of Physics, School of Science and Research Center for Industries of the Future, Westlake University, Hangzhou 310030,  China}
\affiliation{Institute of Natural Sciences, Westlake Institute for Advanced Study, Hangzhou 310024, China}

\author{Bin-Bin Mao}
\affiliation{School of Foundational Education, University of Health and Rehabilitation Sciences, Qingdao 266000, China}

\author{Heng Shen}
\email{hengshen@sxu.edu.cn}
\affiliation{State Key Laboratory of Quantum Optics Technologies and Devices, Institute of Opto-Electronics, Shanxi University, Taiyuan, 030006, China }
\affiliation{Collaborative Innovation Center of Extreme Optics, Shanxi University, Taiyuan, 030006, China }

\author{Zheng Yan}
\email{zhengyan@westlake.edu.cn}
\affiliation{Department of Physics, School of Science and Research Center for Industries of the Future, Westlake University, Hangzhou 310030,  China}
\affiliation{Institute of Natural Sciences, Westlake Institute for Advanced Study, Hangzhou 310024, China}

\begin{abstract}
The entanglement entropy probing novel phases and phase transitions numerically via quantum Monte Carlo has made great achievements in large-scale interacting spin/boson systems. In contrast, the numerical exploration in interacting fermion systems is rare, even though fermion systems attract more attentions in condensed matter. The fundamental restrictions is that the computational cost of fermion quantum Monte Carlo ($\sim \beta N^3$) is much higher than that of spin/boson ($\sim \beta N$). Here, $N$ is the total number of sites and $\beta$ is the inverse temperature or projection length. To tackle this problem, we propose a fermionic quantum Monte Carlo algorithm based on the incremental technique along physical parameters, which greatly improves the efficiency of extracting entanglement entropy. We benchmark the developed algorithm by calculating the scaling behavior of the entanglement entropy in a two-dimensional square lattice Hubbard model. The obtained phase diagram including Fermi surface and Goldstone modes validates the correctness of the algorithm. Remarkably, our method shows the high-efficiency with respect to the existing algorithms, while keeping the high computation precision. We proceed to apply this algorithm to explore the scaling behavior of the entanglement entropy and particularly its derivative at Gross-Neveu criticality. Our results elucidate that such critical behavior can be quantified by the correlation length exponent.
\end{abstract}

\date{\today}
\maketitle

\textcolor{blue}{\it Introduction.--} 
Quantum entanglement, a key non-classical resource in quantum information processing, recently has been found to be one of the fundamental mechanisms of condensed matter physics~\cite{Horodecki2009Quantum,Amico2008entanglement,Laflorencie2016,zeng2019quantum}. In practice, the entanglement entropy (EE) is generally used as a measure of quantum entanglement, especially in many-body physics. While quantum field theory and conformal field theory have difficulties in complex systems or near certain quantum criticalities~\cite{casini2007universal,casini2012positivity,Calabrese2008entangle, Fradkin2006entangle, Nussinov2006, Nussinov2009, JiPRR2019, ji2019categorical, kong2020algebraic, XCWu2020,ding2008block, Tang2020critical,XCWu2021}, numerical methods offer an universal approach to calculate EE, and reveal intrinsic properties beyond local operators, such as the topological order, Goldstone modes, and the value of central charge~\cite{vidal2003entanglement,Calabrese_2004,Eisert2010Colloquium,Fradkin2006entangle}. Recent decade has witnessed significant progress in developing efficient algorithms for large-scale, high-dimensional  interacting systems~\cite{Hastings2010,Humeniuk2012,Grover2013entanglement,JRZhao2020,JRZhao2021,BBChen2022,YCWang2021DQCdisorder,YCWang2021U1,jiangFermion2022,Fermion2023Liu,zyan2021entanglement,JRZhao2022,Liao2023The,Liu2024Disorder,latorre2004,Legeza2006,Chan2008,Ren2012,liu2024measuring,Laurell2023,Li2008entangle,Poilblanc2010entanglement,wu2023classical,liu2023probing,song2023different}.

Among these, the quantum Monte Carlo (QMC) method is by far one of the most promising algorithms for large-scale sign-free systems in two and higher dimensions. Its efficiency is not limited to specific forms of EE, no matter area law or volume law. Although the QMC algorithms of spin/boson systems with an $O(\beta N)$ complexity have been widely leveraged to obtain entanglement information in various phases and phase transition points~\cite{Hastings2010,Humeniuk2012,Grover2013entanglement,luitz2014improving,JRZhao2020,JRZhao2021,wang2024ee,mao2023sampling,li2023relevant,ding2024tracking,song2023deconfined,deng2024diagnosing},  
the QMC algorithms of EE in fermion systems are few because of the algorithmic structure with $O(\beta N^3)$ complexity for determinant calculations. 
Therefore, for a long time, research on the EE of fermionic systems grows slowly. 
However, the main topic of condensed matter are the emergent phenomena in interacting-electron systems, such as high-temperature superconductivity, quantum Hall effect, and twisted bilayer materials, all of which are fermionic. How to extract the entanglement properties of these fermion systems is thus an important but challenging issue. 

The pioneering QMC work for calculating EE in fermionic systems was proposed by Grover, which is based on determinantal QMC (DQMC)~\cite{Grover2013entanglement}, later extended to projection QMC (PQMC)~\cite{d2020entanglement,d2024universal}. Despite theoretical rigorousness, it becomes cumbersome when dealing with situations far from the free fermion limit. 
Therefore, the incremental technique was generalized to fermionic QMC~\cite{d2024universal,Liao2023The,Zhang2024Integral,Liao2023Controllable,pan2023stable}, which has been maturely employed in bosonic QMC~\cite{Hastings2010,luitz2014improving,d2020entanglement,Incremental2024Zhou}. The key spirit is smoothly connecting two far-away distributions by inserting several intermediate distributions, thereby the importance sampling can be realized. Here the two far-away distributions mean the distributions of the partition function and of the targeted observable. We should emphasize that, although the incremental technique has highly improved the precision of the EE data measured by QMC, the virtually introduced intermediate-processes largely increase the computational cost. This results from the fact that in order to hold the importance sampling, the number of intermediate processes needs to be kept as an algebraic growth with system length $L$~\cite{ding2024reweightannealing,Liao2023Controllable}. Therefore, the improvement of computational efficiency for extracting EE in fermion QMC is highly demanded. 

In this Letter, we develop a fermionic QMC algorithm for calculating the EE with high-efficiency and low-computational-cost, offering the possibility for exploring the EE in large-scale interacting-fermion systems. The key idea is to design the specific update flows along a meaningful path in the physical parameter space. In this case, all the intermediate products of the incremental process turns into the EE values of different parameter. 
We apply our algorithm to the two-dimensional square lattice Hubbard model as a benchmark. As expected, we identify the transition between the Fermi surface and the Goldstone modes through the scaling behavior of EE. Importantly, our method generates hundreds of EE values in the parameter space, demonstrating the high efficiency with respect to the existing algorithms. We next employ this algorithm to investigate the scaling behavior of the entanglement entropy and its derivative at the Gross-Neveu critical point. Our findings show that the critical behavior can be quantitatively characterized by the correlation length exponent.

\textcolor{blue}{\it Method.--} 
\label{sec:seci}
We take PQMC as an example to illustrate the mechanism of our algorithm. In fact, this method can be also implemented in DQMC \cite{White1989Numerical,G1986Auxiliary}  generally in the same spirit.
As routinely doing in QMC simulation, we consider the calculation of second R\'enyi EE $S^{(2)}_M$ defined on the subregion $M$ ($\overline{M}$ is the environment) for general interacting fermions. Accordingly, $S^{(2)}_M = -\ln \Tr \rho_M^2$, where $\rho_M$ is the reduced density matrix of subregion $M$. In PQMC regime, $S^{(2)}_M$ can be formularized as the ratio of two partition functions, 
\begin{equation}
  S^{(2)}_M = -\mathrm{ln}\frac{Z_M^{(2)}}{Z^2},
  \label{eq:1}
\end{equation}
where $Z_M^{(2)} = \sum_{s_1,s_2} W_{s_1} W_{s_2} \det g_{M,s_1,s_2} $, $Z = \sum_{s_1}W_{s_1}$ with the auxiliary fields labeled $s_1$ and $s_2$~\footnote{$Z^2$ is the square of the partition function of the whole system, where in practice one could only simulate $Z$ and then square.}, while $\det$ denotes determinant calculation. $g_{M,s_1,s_2} =  G_{M,s_1} G_{M,s_2} + (\mathbbm{1} - G_{M,s_1} )(\mathbbm{1} - G_{M,s_2} ) $ is referred to as the Grover matrix, which is decided by the Green function matrix $G$ for both $s_1,s_2$ \cite{Grover2013entanglement}. $W_{s_1}$ represents the standard configuration weight of $s_1$ in QMC. 

Notice that direct calculating the ratio between ${Z_M^{(2)}}$ and ${Z^2}$ to obtain the EE by Eq.\eqref{eq:1} is straightforward and theoretically rigorous, however, the ratio is exponentially small as the system size increases, resulting in a poor sampling efficiency. To overcome this difficulty, the incremental technique has been introduced by dividing the small value of the overlap into the product of several larger values \cite{DEmidio2020,JRZhao2021,JRZhao2022,d2024universal,Zhang2024Integral}. 
A feasible way is to estimate the EE value at a certain parameter point from that at its nearby point. This is so-called the ``reweight" approach. The ratio of two close partition functions $\mathcal{Z}$ can be measured through the averaged ratio between two related weights~\cite{ding2024reweightannealing},
\begin{equation}
  \frac{\mathcal{Z}(f_n)}{\mathcal{Z}(f_{n-1})} 
  = \bigg\langle \frac{\mathcal{W}_{s_1,s_2}(f_{n})}{\mathcal{W}_{s_1,s_2}(f_{n-1})} \bigg\rangle_{f_{n-1}}.
  \label{eq:ave}
\end{equation}
Here $\mathcal{Z}(f)$ and $\mathcal{W}_{s_1,s_2}(f)$ represent the general type of partition function (either $Z$ or $Z_M^{(2)}$) and weight at parameter set $f$. 
In the realistic simulation, the result of the reweighting is good only if the two parameter points $f$ and $f_0$ are close enough, i.e., the ratio is closed to 1 \cite{ding2024reweightannealing,dai2024residual,neal2001annealed}. Otherwise, we naturally insert several intermediate points to split the reweighting process, 
\begin{equation}
  \frac{\mathcal{Z}(f_n)} {\mathcal{Z}(f_0)}= \frac{\mathcal{Z}(f_n)}{\mathcal{Z}(f_{n-1})} \times ... \times \frac{\mathcal{Z}(f_1)}{\mathcal{Z}(f_0)},
  \label{eq:expansion}
\end{equation}
where $n$ is the number of slices. By specifying $\mathcal{Z}$ to $Z^{(2)}_M$ and $Z$ in the above formula, the numerator ${Z_M^{(2)}}$ and denominator $Z^2$ in Eq.\eqref{eq:1} can be obtained respectively. 
Moreover, each term of partition function ratio in Eq.\eqref{eq:expansion} can be computed parallelly. On the other hand, a known reference $\frac{Z_M^{(2)}(f_0)}{Z^2(f_0)}$ is required to fix the value of $\frac{Z_M^{(2)}(f_n)}{Z^2(f_n)}$. In general, we prefer to choose a product state with $\frac{Z_M^{(2)}(f_0)}{Z^2(f_0)}=1$.

Here, we outline the key points of the algorithm for simulating Eq.\eqref{eq:ave} in interacting fermionic systems. Different from conventional QMC algorithm, the observable sampling $\langle \frac{\mathcal{W}_{s_1,s_2}(f_{n})}{\mathcal{W}_{s_1,s_2}(f_{n-1})} \rangle$ requires two sets of computer memory space labeled as $f_n$ and $f_{n-1}$. Each memory stores the singular value decomposition (SVD) matrix structure on each time slice, the equal-time Green function and other intermediate variables of the PQMC program. Importantly, two memory spaces share the \textit{same} copy of the auxiliary field $s_1,s_2$, which is updated only according to the weight of $f_{n-1}$. The other essential element is to keep the weight $\mathcal{W}_{s_1,s_2}(f_n)$ and $\mathcal{W}_{s_1,s_2}(f_{n-1})$ as a global variable to simplify the observable calculation. Finally, regarding the choice of the parameters, we typically fix several parameters in a series of parallel PQMC programs, such as the Trotter decomposition interval $\Delta\tau$, the projection length $\beta$ and the system size $L$. Hence the number of the imaginary time slices, which is determined by $\beta$ and $\Delta \tau$, also maintains as a constant (See SM for more details). 

\textcolor{blue}{\it Algorithm benchmark.--} 
To give a benchmark of the algorithm, we choose the 2D square lattice Hubbard mode  as an example, and perform PQMC simulation parallelly with half of the whole system as entangled region (See SM for the details).  In order to set an appropriate path in the parameter space, we set the hopping strength $t = 1$ as an unit, and only vary the interaction strength $U$ as the path in parameter space for EE calculation. In practice, we choose the adjacent parameter sets to fulfill the condition that the ratio between two closest partition functions is a moderate constant, e.g., ${\mathcal{Z}(f_n)}/{\mathcal{Z}(f_{n-1})}\in [0.1,10]$ (See SM for the discussion on the choice of the intervals). With this strategy, the importance sampling is maintained and segmentation grows algebraically with size.





\begin{figure}[t]
        \centering
        \includegraphics[width=\columnwidth]{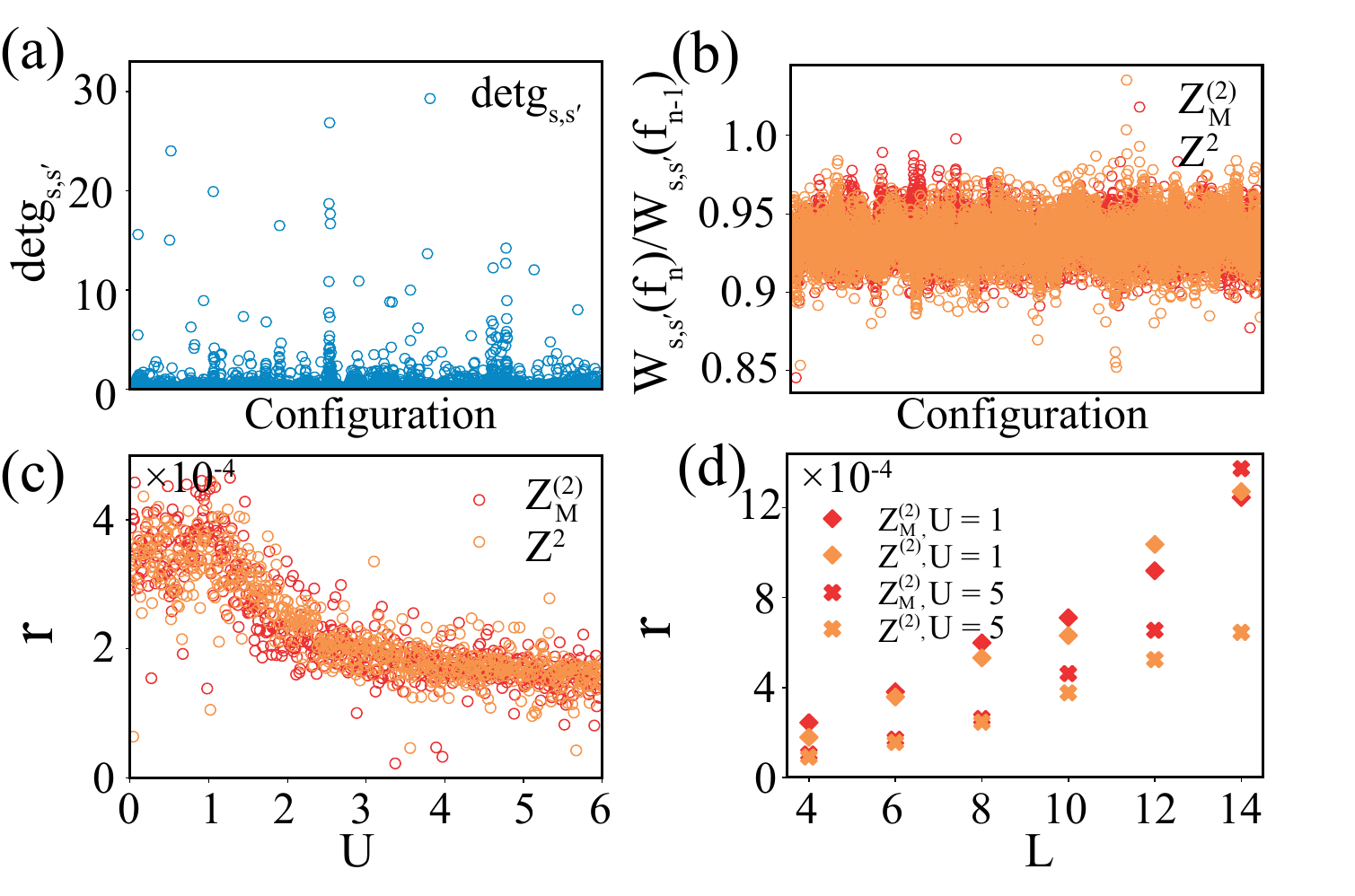}
  \caption{ Comparison in the observable distribution between Grover's methods in (a) and new algorithm in (b). The parameter is chosen as $L= \beta =4$. $U=8$ for Grover's method, and the parameter interval $(U, U+\Delta U) = ( 8, 8.01 )$ for the new algorithm. (c) The defined parameter $r$ (ratio between the sample standard deviation and sample mean value) as a function of $U$. Here, $L=\beta=4$. The value of $r$ is at the level of $10^{-4}$. $r$ for both two observables have little difference. (d) $r$ as a function of $L$. At free fermion limit, the system size-dependence of both two observables are almost same. While far from free fermion limit, $r$ of the observables which contains Grover matrix increases faster than others with the system size.} 
  \label{fig:1}
\end{figure}

We start by demonstrating the stability and convergence of the algorithm. In fact, it is a problem raised in the Grover's algorithm~\cite{Grover2013entanglement} that the few measurements deviate from the most values, as shown in Fig.~\ref{fig:1}(a), which is absent in our algorithm. Because such exceptional observed values are converted to the ratio of two exceptional values of the same configurations at adjacent Hamiltonian parameters, according to Eq.\eqref{eq:ave}, and we numerically observe that the ratio no longer deviates from the majority, as depicted in Fig.~\ref{fig:1}(b). Additionally, the distribution of the observables is narrow. Therefore, all above observations indicate high controllability and validity for the errorbar in a single calculation.
To evaluate the distribution quantitively, we define one parameter, named $r$, in the computation of $Z$ and $Z^{(2)}_M$ to characterize the convergence of the observable. It is the ratio between the sample standard deviation and sample mean value. 
We find that $r$ varies little with interaction strength $U$ [Fig.~\ref{fig:1}(c)] at small system size, at least in the same order of magnitude. Fig.~\ref{fig:1}(d) displays the size-dependence of $r$ at small ($U=1$) and large ($U=5$) interaction strength. In particular, at large $U$, we observe that $r$ associated with the $Z_M^{(2)}$ polynomially depends on the system size $L$, whose value is relatively small at small system size [Fig.~\ref{fig:1}(d)]. This suggests that more computation cost is required in the large system to promise the same precision as that with small system size, Nonetheless, it is still within our reach for computation resource. 

\begin{figure}[t]
    \centering
    \includegraphics[width=\columnwidth]{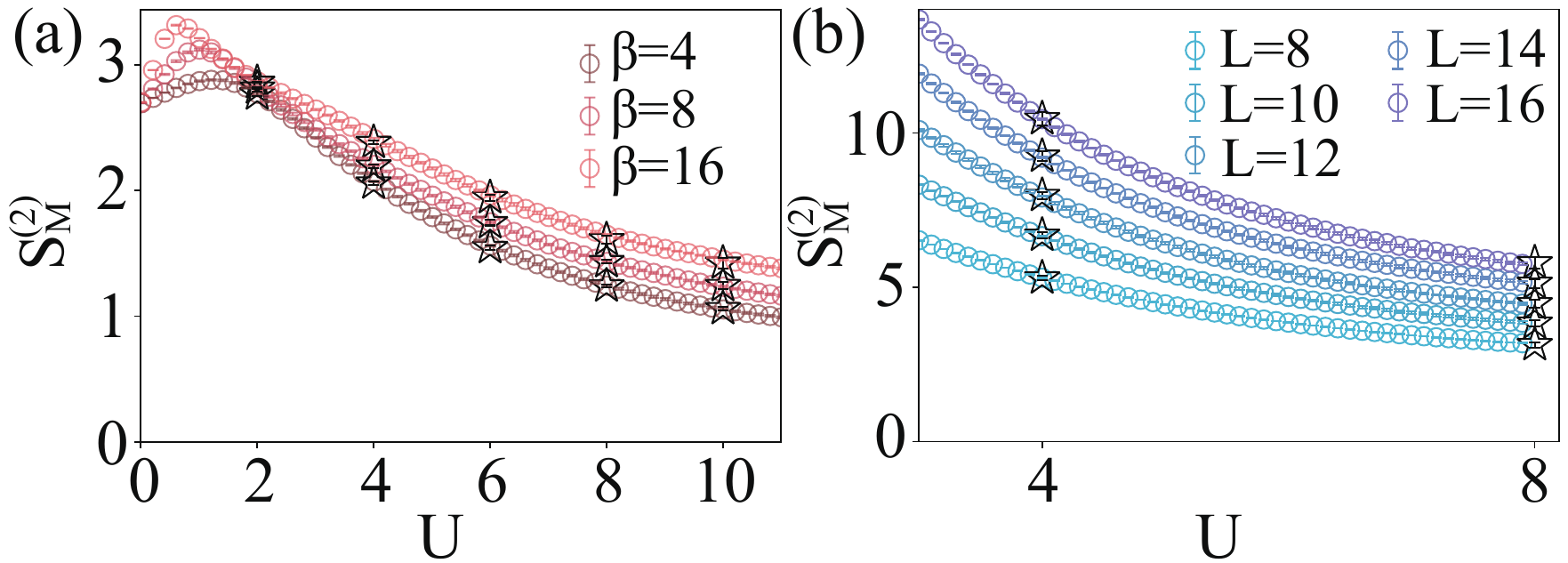}
    \caption{Comparison in EE results between the new algorithm marked by colored line, and previous methods marked by corresponding colored dots. (a) $S_M^{(2)}$ as a function of $U$ for various projection length $\beta$ given $L=4$. The data of pentagram comes from Grover's method~\cite{Grover2013entanglement}. (b) $S_M^{(2)}$ as a function of $U$ for various system size $L$ under $L=\beta$. The data of pentagram comes from D'Emidio's method~\cite{d2024universal}, and $U=8$ data is adapted to Ref.~\cite{pan2023stable}. Note all computations are obtained without TBC.} 
  \label{fig:2}
\end{figure}

We further verify the accuracy and the correctness of EE results by comparing with the EE obtained through other methods. Firstly at small system size $L=4$ [Fig.~\ref{fig:2}(a)], the EE via our algorithm (marked by circle) is consistent with that obtained by the Grover's methods (marked by pentagram). Nevertheless, Grover's method becomes unfaithful at large system size. Therefore, it is adequate to consider the equilibrium algorithm recently proposed by D'Emidio~\cite{d2024universal} as a benchmark. Here, we refer to the data in Ref.~\cite{pan2023stable} by D'Emidio's method under the strong interaction $U=4$ and 8, which is marked by pentagram in Fig.~\ref{fig:2}(b). The result is in good agreement up to the system size $L=16$. All the above comparisons indicate the high data quality and correctness of our algorithm. We would like to emphasize that Fig.~\ref{fig:2} reveals the advance of our method -- given the similar computation time, other methods obtain one data point while we gain a data curve. This attributes the fact that the incremental process of our method is along a real physical parameter path.

We proceed to uncover the physics behind the square lattice Hubbard model by scanning the EE behavior in the parameter space. To perform a good convergence to the ground state, we exploit the twisted boundary condition (TBC), acting on the choice of the trial wavefunction (See SM for the details). As is well-known, the two dimensional square lattice Hubbard model holds a metal-insulator transition. In two phases, the associate scaling behavior of EE versus the length of subregion is distinct. 
The general form is written as,
\begin{equation}
  S^{(2)}_M =  A L \ln L + a L + b \ln L + c.
  \label{eq:scaling1}
\end{equation}

In the absence of $U$, the model behaves as the metal with a Fermi surface (FS), whose scaling behavior of EE is dominated by the characteristic leading term $L \ln L$.
The leading term coefficient $A$ is determined by the shape of both the FS and subregion $M$, expressed by the Widom-Sobelev formula~\cite{Gioev2006Entanglement,Leschke2014Scaling}\footnote{In the free fermion limit of our model, the $A=0.5$ is known (See Supplementary Materials for details)}. 

When adding positive $U$, the gap gradually opens and the system turns into an insulator. In such a phase, the coefficient $A$ in Eq.\eqref{eq:scaling1} vanishes and the associated EE shows an area law. Deep in the insulating phase, the coefficient of the $\ln L$ term $b$ equals to $N_G/2$ under a bipartite corner-less cutting, where the $N_G=2$ is the number of Goldstone modes in N\'eel order \cite{Metlitski2011}. 
In our simulation, the subregion is chosen as the rectangle shape to exhibit the Goldstone mode. 


\begin{figure}[htp!]
    \centering
    \includegraphics[width=0.9\columnwidth]{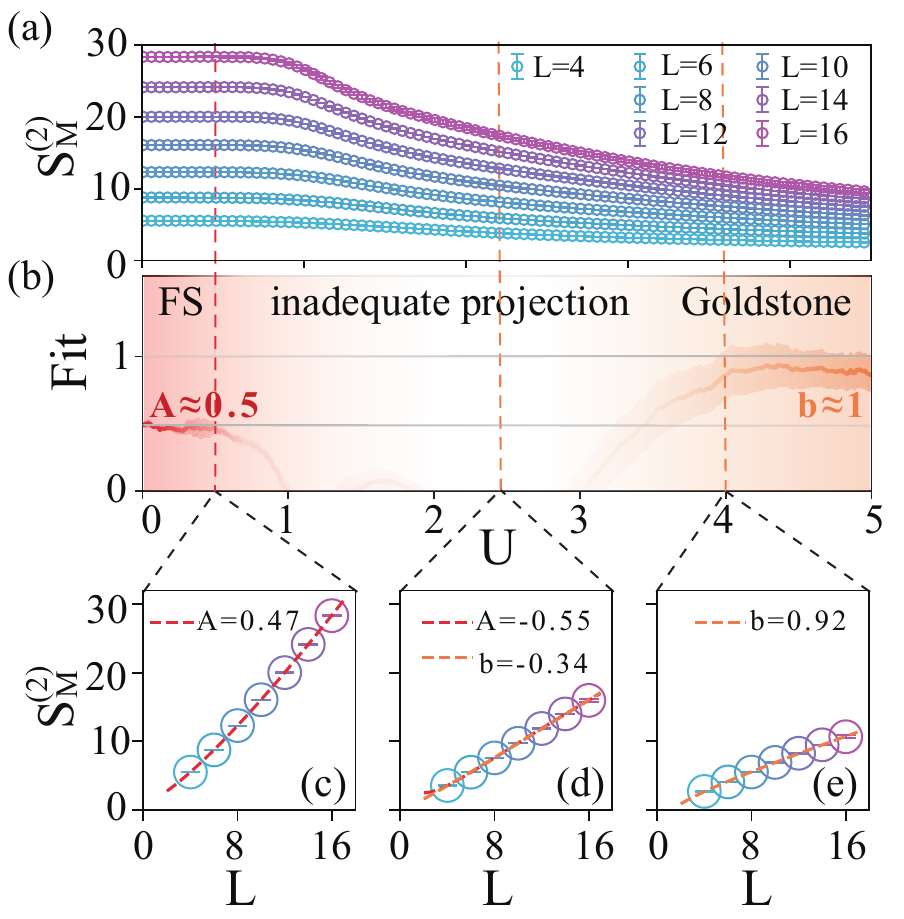}
    \caption{(a) EE obtained from the new algorithm as a function of $U$ for various $L$ up to 16. Here $\beta = L$ for all curves. (b) Fitting result for the scaling function of EE at various $U$. At small $U$, we fits EE with Eq.\eqref{eq:scaling1}, where the curve of leading term coefficient $A$ is colored by red. At large $U$, we fits EE with Eq.\eqref{eq:scaling1} given $A=0$, where the curve of universal coefficient $f$ is colored by yellow. The white region represents the intermediate parameter interval, where both function forms are inadequate to give the scaling description. (c) Scaling behavior at $U=0.5, 2.5$ and $4$. The red and yellow curves are fitting results by Eq.\eqref{eq:scaling1}. Note we use TBC for all computations.} 
  \label{fig:3}
\end{figure}

All the physics are supported by the numerical results of EE [Fig.~\ref{fig:3}(a)] and universal coefficients $A$ and $b$ extracted by fitting with Eq.\eqref{eq:scaling1}[Fig.~\ref{fig:3}(b)]. In particular, Fig.~\ref{fig:3}(c)-(e) display the distinct scaling behaviors of EE at three different interaction strengths. Firstly, under the free fermion limit ($U=0$), $L\ln L$ behavior manifests clearly, and fit result of $A$ is close to the $0.5$, i.e., analytic solution in the thermodynamic limit ~\cite{Leschke2014Scaling,Gioev2006Entanglement} (See SM for the derivation). As increasing $U$, the obtained EE monotonously decreases, as shown in [Fig.~\ref{fig:3}(a)], which is valid since the system becomes more insulating. In addition, it is found that at $U \gtrsim 4$, the scaling behavior is in accordance with Eq.\eqref{eq:scaling1} ($A=0$, area law), and $b \sim 1$ remains almost unchanged. The deviation of $b$ from 1 attributes to the strong finite size effect of the effective Heisenberg model in the large $U$ limit ~\cite{deng2023improved}, and actually it is normal that $b<1$ in numerical simulations. 

However, we find that the goodness-of-fit becomes worse near the free fermion limit, due to the inadequate projection, i.e. the projection length $\beta$ is not large enough. The problem could be solved by choosing proper trial wavefunctions to a certain degree (See SM for the details). Nevertheless, our algorithm successfully displays such features at insufficient projection length. 

\textcolor{blue}{\it Scaling behavior of EE around Gross-Neveu criticality.--} 
Equipped with this algorithm, we explore the physics of scaling behavior of EE around Gross-Neveu criticality in the Hubbard model defined on the $\pi$-flux square lattice. Such model exhibits the Gross-Neveu transition from the semi-metal phase at small $U$ to Mott insulator phase at large $U$\cite{Parisen2015Fermionic,Universal2016Otsuka}. And the EE behavior is formalized as, 
\begin{equation}
  S^{(2)}_M =  a L + b \ln L + c,
  \label{eq:scaling2}
\end{equation}
with no FS contribution. At semi-metal phase, the corner of entangled region contributes to the logarithmic term coefficient $b$, as discussed in literature~\cite{Universal2016Helmes,Fermion2023Liu,Liu2024Disorder}. However, the critical behavior of the coefficient $a$ remains elusive. 

In fact, the $O(N)$ criticality also follows the EE behavior of Eq. \eqref{eq:scaling2} ~\cite{Metlitski2009,Helmes2014,wang2024probing}, where the constant term $c$ is a characteristic for probing the $O(N)$ criticality. Strikingly, the coefficient $a$ associated with the leading term satisfies
$a(g)-a(g_c)\sim |g-g_{c}|^{\nu}$~\cite{Metlitski2009} according to the expansion of field theory, where $g$ is the tuning parameter of Hamiltonian, and $g_c$ is the critical point. $\nu$ is the correlation length critical exponents, whose value is depended on the universality class of the phase transition. We should emphasize that the EE behavior of $O(N)$ criticality is derivated from the small $\epsilon$~\footnote{$\epsilon=3-d$, d is spatial dimension} and large $N$  expansion for the quantum $O(N)$ model, and this is not applicable to the Gross-Neveu phase transition. It is thus natural
to raise the question as whether one can extend the scaling relation between $a$ and $\nu$ to the Gross-Neveu criticality.

\begin{figure}[htp!]
    \centering
    \includegraphics[width=\columnwidth]{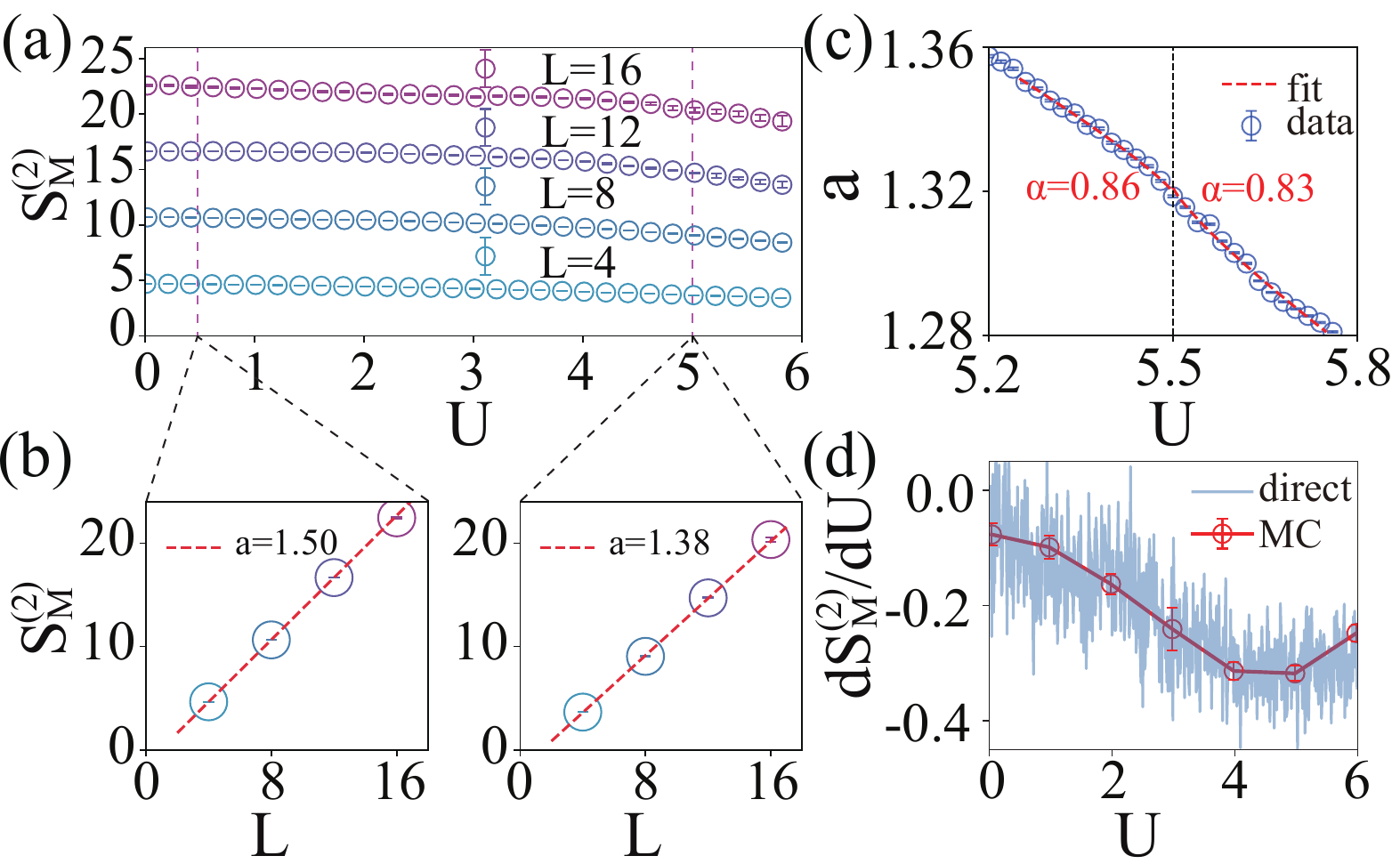}
    \caption{EE and its derivative in $\pi$-flux model. (a) EE as a function of $U$ for various $L$ up to $L=16$. Here $\beta = L$ for all curves. (b) Fitting results for $U=0.5$ and $U=5.0$. (c) Fitting results of EE near quantum criticality. The blue data points are the fitting of coefficient $a$ for the linear function in Eq.\eqref{eq:scaling2}, note $b=0$ for smooth boundary. We fix $U_c = 5.5$ and fit $a(U)$ for the power law behavior from both sides, which provides almost unified results of $\alpha$. (d) The comparison of the derivatives between the direct way and MC calculation. For the direct approach, we use the difference of the adjacent EE values to obtain its derivative. }
  \label{fig:4}
\end{figure}

Here, we perform the simulation with the half of the lattice as the entangled region without corner (See SM for the details) to ensure $b=0$. Fig.\ref{fig:4}(a) shows the $U$-dependence behavior of EE obtained by our algorithm, where we extract the EE results at certain $U$ points, which show a clear linear size-dependence behavior [Fig.\ref{fig:4}(b)].
In Fig.\ref{fig:4}(c), the power law fitting near the known quantum critical point $U_c \sim 5.5$ is plotted, where we observe a nearly linear fitting with the power $\alpha \sim 0.85$, which quantitatively agrees with the results in previous MC simulation~\cite{Fermionic2015Parisen}, where they give $\nu = 0.85(4)$ for the largest system size. We therefore conclude that our findings could support the conclusion that the $O(N)$ phase transition also applies to the Gross-Neveu criticality.  



As a further analysis, we also put forward a method to calculate the EE derivative in high precision. In particular, we could just assign $\frac{\partial \mathcal{W}_{s_1,s_2}}{\partial f}$ as observables under the original program implementation (See SM for the details). In Fig.\ref{fig:4}(d), we compare the above method with the direct derivatives of the EE data in (a), and find the former has smaller errorbar. Finally we also note the biggest advantage for derivatives calculation is that no incremental method is required from a known starting points. In other words, we could directly explore the region of interest in the phase diagram, for example, near quantum criticality, by using the EE derivative instead of EE. 

{\it Summary and outlook.--} 
We report an efficient fermionic QMC algorithm with amount of EE data in once simulation to fix the difficulty of the heavy computational cost of the EE calculation in large-scale interacting fermion systems. 
This is distinguished from the existing methods, where one can get only one data in a single implementation. Our algorithm provides the opportunity to scan EE for exploring its relation against the Hamiltonian parameters. The intrinsic physics in square lattice Hubbard model has been revealed via the EE by our method, such as the FS in $U\rightarrow 0$ limit and Goldstone modes in large $U$ limit. Furthermore, to manifest the advantage of the new algorithm, we apply it to the EE study of the $\pi$-flux Hubbard model, where the associated phase transition is at finite $U$ with Gross-Neveu universality class. Numerically, we  find the scaling behavior of EE seems to be determined by the correlation length exponent, similar to the $O(N)$ phase transition. Meanwhile, we also design the EE derivative algorithm to obtain high-precision derivative data. Given the fact that the highly entangled matter in large-scale and high-dimensional systems plays an essential role in condensed matter and statistic physics, significant efforts has been recently put in developing the numerical methods for spin/boson system, yet the counterpart for fermion systems is rare, even though fermion systems are of great interest in condensed matter. Our method thus offers an ubiquitous tool for exploring the physics in the interacting fermion systems.

{\it Acknowledgment.--} We thank the helpful discussion with Wei Zhu, Yao Zhou, Peng Ye, Zi Hong Liu and Xiaofan Luo. This work is supported by National Natural Science Foundation of China (Project No.\ 12404275 and 12222409), and the fundamental research program of Shanxi province (Project No.\ 202403021212015). Z.W. and Z.Y. acknowledge the China Postdoctoral Science Foundation under Grants No.2024M752898, the Scientific Research Project (No.WU2024B027) and the start-up funding of the Westlake University. G.P. acknowledges the Würzburg-Dresden Cluster
of Excellence on Complexity and Topology in Quantum
Matter - ct.qmat (EXC 2147, Project No. 390858490). H. S. acknowledges the Royal Society Newton International Fellowship Alumni follow-on funding (AL201024) of UK. The authors thank the high-performance computing center of Westlake University and the Beijng PARATERA Tech Co.,Ltd. for providing HPC resources. 

\bibliography{ee}

\clearpage

\setcounter{equation}{0}
\setcounter{figure}{0}
\renewcommand{\theequation}{S\arabic{equation}}
\renewcommand{\thefigure}{S\arabic{figure}}
\setcounter{page}{1}
\linespread{1.05}
\begin{widetext}
    
\centerline{\bf\Large Supplemental Material} 

\section{Sketch map of the Hubbard Model}
In the main text, we have introduced the Hubbard model defined on squared lattice and $\pi$-flux lattice. Below we display the sketch map of the lattice structure, along with the choice of the entangled region.

\begin{figure}[h]   
    \includegraphics[width=0.5\textwidth]{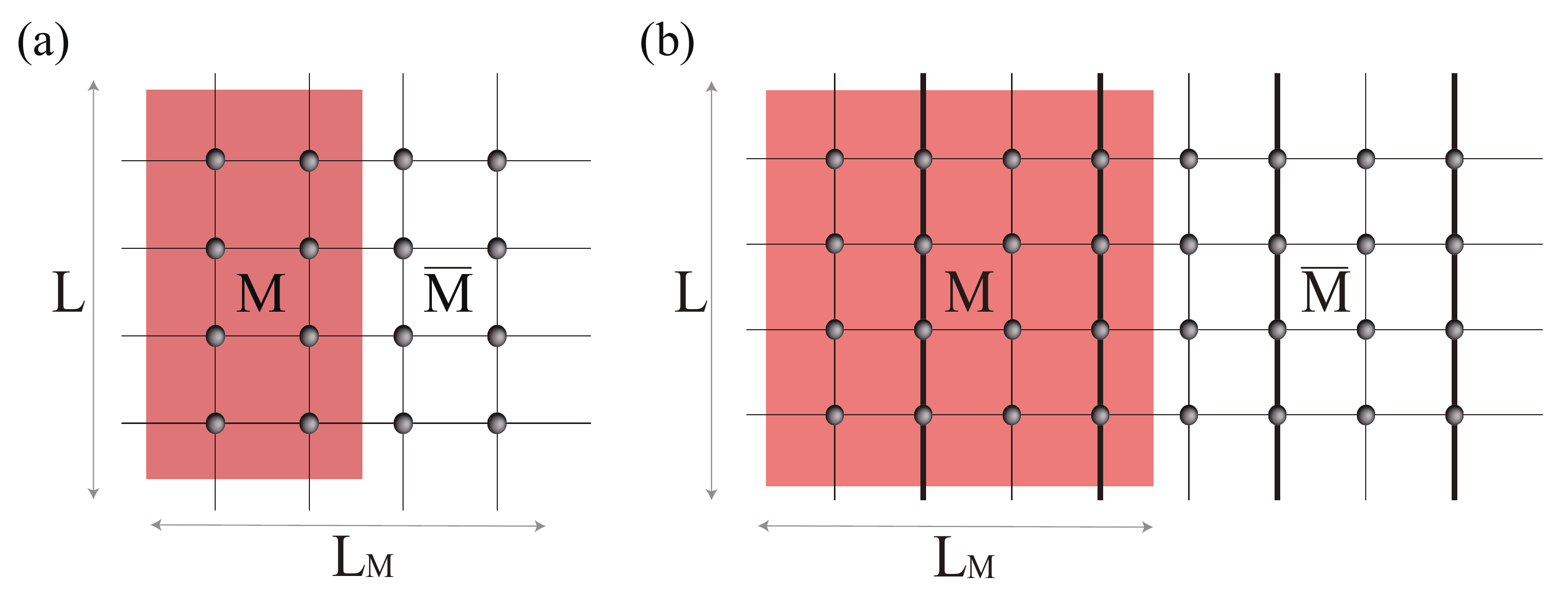}
     \caption{Sketch map for the square lattice and $\pi$-flux lattice with period boundary condition. The entangled subregion is denoted as $M$, with its complementary part $\bar M$. Each entangled region has a rectangle shape colored by red, and is half of the whole system. (a) Square lattice with size $L \times L$. $M$ has size $L \times L_M$. For square lattice $L_M = L/2$. The bonds between lattice sites represent the uniform hopping amplitude $-t$. (b) $\pi$-flux with size $L \times 2L$. $M$ has size $L \times L_M$. Here $L_M = L$. The thin and thick bond represents hopping amplitude $-t$ and $t$.}
\end{figure}

\section{Algorithm details}

In this section, we provide detailed implementation of the new algorithm and its complexity analysis. In Fig.\ref{fig:algo}, we show the simple flow chart of the algorithm.

\begin{figure}[h]   
    \includegraphics[width=0.5\textwidth]{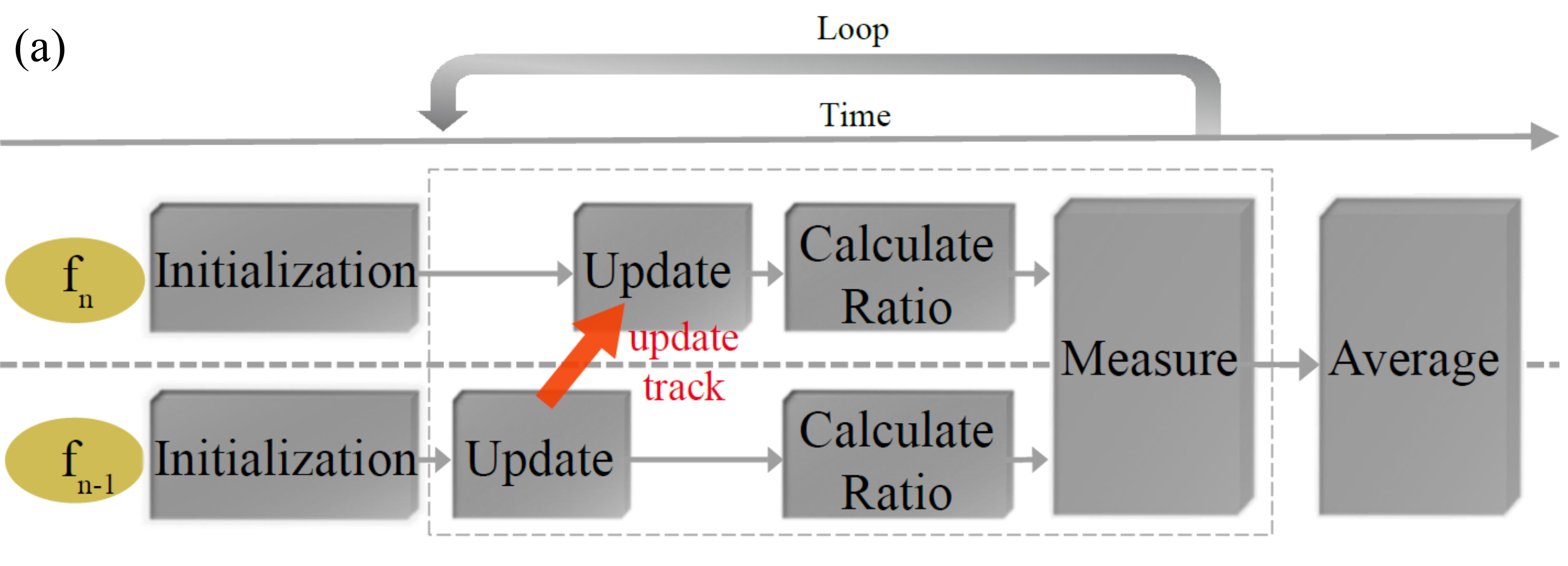}
     \caption{Algorithm flow diagram for calculating $\mathcal{Z}(f_n)/ \mathcal{Z}(f_{n-1})$. The upper and lower part separated by the dashed line represents two identical memory space for two parameter sets $f_n$ and $f_{n-1}$, respectively. Note the update results are the lattice site index and whether these updates occur, according to the ratio of $f_{n-1}$. }
     \label{fig:algo}
\end{figure}

Firstly, we initialize the program variables subject to both parameter sets $f_{n-1}$ and $f_n$ without any update regime. In this process, we exactly calculate the initial weight of $\mathcal{W}_{s_1,s_2}(f_n)$ and $\mathcal{W}_{s_1,s_2}(f_{n-1})$ for random auxiliary field $s_1,s_2$ and keep the weight. We also obtain Single Value Decomposition matrix for both parameter sets, which is used for numerical stability operation and the weight calculation. After initialization, we enter the cycle of update and measurement, indicated as grey dashed box in Fig.1 (a) in the main text. In each cycle, we first perform a general update step as the origin PQMC algorithm for the whole space-time lattice, according to the weight subject to $f_{n-1}$. In practice, we adopt single-site update, since more simplifications are employed to give a faster computation of the weight and the update ratio. During this process, the updated results marked by red text in Fig.1 (a) in the main text are stored. Such results contain the update sequence of the lattice site index, for example $\{ 16, 8, 25, 23\cdots\}$ and whether the auxiliary field configuration defined on the corresponding lattice is changed, for example $\{ 0, 1, 1, 0\cdots\}$, where 0(1) denotes flipped(not flipped). Subsequently, we update program variables subject to $f_n$ directly according to the stored update results, without any probabilistic criteria. 
Notice that in this step we use the weight before the update in combination with the update result of $f_{n-1}$ to obtain the updated weight by lower cost, instead of recalculation (See Fast update procedure sections for details). 
Since both weights subject to $f_{n-1}$ and $f_n$ are already calculated, we conduct a measurement after update process. By repeating this cycle many times, we finally gather the whole measurement data and then take the averaged value as the general Monte Carlo algorithm.

We proceed to discuss the complexity of this algorithm. Once equipped with the above technique, the partition function can be calculated along with the path we chosen in the parameters space. Here, a difficulty arises from the calculation of the determinant of the Grover matrix $\det g_{M,s_1,s_2}$. To effectively manage the Grover matrix, we adopt the algorithm in Ref.~\cite{d2024universal}, and always store the $g_{M,s_1,s_2}^{-1}$ in memory for the subsequent operations. Specifically, when applying single site update, the updated inversed Grover matrix is related to the matrix elements before the update. Hence, $O(N_M^2)$ complexity is achieved instead of recalculation with $O( N_M^3)$  complexity in each update step for Grover matrix. Here, $N_M$ is the number of the site in the subregion $M$. Even if $N_M$ scales linearly with the number of the whole system sites $N$, the total computation complexity is $O(\beta N^3)$, which also takes account of the complexity of other update process for the auxiliary field. This complexity is on the same scale of the normal PQMC algorithm. 

For the derivatives calculation, the main idea originates from the simple expression $\frac{\partial \ln \mathcal{Z}}{\partial f} = \frac{1}{\mathcal{Z}} \frac{\partial \mathcal{Z}}{\partial f}$, where $\mathcal{Z} = \sum_{s_1,s_2} \mathcal{W}_{s_1,s_2}$. The right-hand side is easily to realize through MC sampling, by assigning $\frac{\partial \mathcal{W}_{s_1,s_2}}{\partial f}$ to be the observable and updating according to the configuration weight $\mathcal{W}_{s_1,s_2}$. To be specific, based on the algorithm implementation that $\mathcal{W}_{s_1,s_2}$ is kept during the whole program, we are able to calculate its difference form $\frac{\delta \mathcal{W}_{s_1,s_2}}{\delta f}$, where $\delta \mathcal{W}_{s_1,s_2}(f) = \mathcal{W}_{s_1,s_2}(f + \frac{\delta f}{2}) - \mathcal{W}_{s_1,s_2}(f - \frac{\delta f}{2})$ with the error $O(\frac{\delta f^2}{4})$. The algorithm structure fits perfectly with derivative calculation. In practice, $\delta U$ is chosen to be a tiny value to avoid the error (We choose $\delta U = 0.001$ for Fig. 4 in the main text). 

\section{Fast update procedure}
Here, we provide detailed description of the fast update procedure. In PQMC, the calculation of partition function follows,
\begin{equation}
  Z = \left\langle  \Psi_T\left|e^{-2 \beta H}\right| \Psi_T\right\rangle = C^m \sum_{s} \operatorname{det}\left[P^{\dagger} B_s(2 \beta, 0) P\right].  
\end{equation}
And the weight is expressed as,
\begin{equation}
  W_s = C^m \operatorname{det}\left[P^{\dagger} B_s(2 \beta, 0) P\right],
  \label{eq:S1}
\end{equation}
where $| \Psi_T\rangle$ is the trial wavefunction, whose information is encoded in matrix $P$. $B$ matrix is determined by the Hamiltonian.  Note $C^m$ is omitted in the simulation, because the observables is the ratio of two partition functions, defined as $\langle \frac{\mathcal{W}_{s_1,s_2}(f_{n})}{\mathcal{W}_{s_1,s_2}(f_{n-1})} \rangle$ in the main text. The simplest way to obtain weight after the update $W_{s^\prime}$ is recalculating Eq.\eqref{eq:S1}, which needs SVD matrix and cost $O(N^3)$ complexity. In practice, we adopt single site update, where Sherman-Morrision methods reduce the complexity to $O(N^2)$ by calculating the update ratio $R=\frac{\operatorname{det}\left[P^{\dagger} B_{s^{\prime}}(2 \beta, 0) P\right]}{\operatorname{det}\left[P^{\dagger} B_s(2 \beta, 0) P\right]}$. In addition, considering the observables is more complex than original PQMC, one operation is design to always keep $W_s$ in the memory space for the observable calculation.
That requires only one exact calculation of $W_s$ at the beginning, and then repeated update $W_s$ to $W_{s^\prime}$ using the calculated ratio $R$. Then the total complexity remains $O(\beta N^3)$, in the same order as the original PQMC. In a word, the fast update process naturally offers to update for the observables, in which case we call this algorithm "passing the weight".

Such an idea can be also realized in the presence of $\det g_{s_1,s_2}$ in the weight. We have,
\begin{equation}
  \begin{aligned}
  Z^{(2)}_M &= \sum_{s_1,s_2} W_{M,s_1,s_2} = \sum_{s} W_{M,s},\\
  W_{M,s} &=  W_{s} W_{s^\prime} \det g_{M,s_1,s_2},
  \end{aligned}
\end{equation}
where the weight $W_{M,s}$ contains two parts, the former is identical to the weight $W_{s}$, and the latter is the determinant of the Grover matrix. In the single site update regime, we use methods proposed by D'Emidio (See Ref.\cite{d2024universal} for details), saving $g_{M,s_1,s_2}^{-1}$ for calculation. Then the complexity for calculating the ratio $R = \frac{\operatorname{det} g_M^{s_1^\prime, s_2}}{\operatorname{det} g_M^{s_1, s_2}}$ ( or $R = \frac{\operatorname{det} g_M^{s_1, s_2^\prime}}{\operatorname{det} g_M^{s_1, s_2}}$ ) scales with $O(N_M^2)$, which equals to $O(N^2)$, in our case $N_M=N/2$. Such process could avoid recalculating the determinant at each measurement to obtain the ratio. Therefore, the total complexity is still controlled in $O(\beta N^3)$. 

Unfortunately, the above method could be problematic due to the passing process. We calculated the exact value of the determinant before and after the single update process and compared it with the ratio. We numerically find the simplified computing method for Grover matrix ratio $R$ may sometimes not be exact. Such inaccuracy could be a negligible effect on the update process, since it only slightly change the update probability. However, the inaccurate $R$ has relatively serious influence on the observable calculations, i.e. the updated weight $W_{M,s^\prime}$ obtained by $W_{M,s}$ and $R$. What is even worse, the error could accumulate and result in completely incorrect results.

To avoid this, we recalculate the determinant at the end of each sweep of space-time sites and conduct it as the exact value of the weight, which contains $\frac{\beta N}{\Delta \tau}$ single updates. The process is similar to the numerical stabilization in auxiliary field QMC. We numerically find the error between recalculation and passing weight process are smaller than $10^{-4}$, in which case, the calculated EE is in accordance with the results from the previous method within the errorbar ( See Fig.2 in the main text ). One can speculate the frequency for doing stabilization depends on the number of the updates, namely, large system size and $\beta$ require more stabilizations. Under such condition, the operation should be adjusted according to the parameters to control the passing error. 

\section{Possible promotion of the new algorithm efficiency for incremental methods}

In the section, we provide the expatiation of the incremental method, and an quantitatively analysis for the degree of the efficiency promotion by dividing the parameter interval. 

The idea of incremental method in MC algorithm is put forward for managing the problematic sampling process, i.e. non-important sampling. A typical manner is to transform the following ratio, 
\begin{equation}
    \frac{Z_M^{(2)}}{Z^2}=\frac{\sum_{s_1,s_2} W_{s} W_{s^\prime} \det g_{M,s_1,s_2} }{\sum_{s_1,s_2}W_{s_1} W_{s_2}}
\end{equation}
into,
\begin{equation}
     \frac{Z_M^{(2)}}{Z^2} = \frac{\sum_{s_1,s_2} W_{s_1} W_{s_2} (\det g_{M,s_1,s_2})^\delta }{\sum_{s_1,s_2}W_{s_1} W_{s_2}}\times  \frac{\sum_{s_1,s_2} W_{s_1} W_{s_2} (\det g_{M,s_1,s_2})^{2\delta} }{\sum_{s_1,s_2}W_{s_1} W_{s_2} (\det g_{M,s_1,s_2})^\delta}\times \cdots \times \frac{\sum_{s_1,s_2} W_{s_1} W_{s_2} (\det g_{M,s_1,s_2})^{n\delta} }{\sum_{s_1,s_2}W_{s_1} W_{s_2}(\det g_{M,s_1,s_2})^{(n-1)\delta}}
\end{equation}
where $\delta=1/n$ and $n$ is a large number to ensure each divided ratio is not too small. In this way, the precision of EE has been improved. However, the intermediate ratios in present methods are unmeaning and consume a lot of computational resources. 
In our new algorithm, we intuitively set the incremental process along a real physical parameter path in this algorithm. All the intermediate products are thus the EE values at different parameters points. As a consequence, the efficiency has been greatly improved through taking advantage of the incremental process. This method produces more EE values than conventional incremental schemes, whose amount increases polynomially as a function of the system size. 

At the very beginning, we emphasize that the analysis is only valid on condition that the exception value problem is not serious, in other words, $\Delta U$ is small enough. Suppose $\mathcal{Z}(f)/\mathcal{Z}(f_0)=1/y$, where $y$ is much bigger than 1. If one divides $[f_0,f]$ into $n$ subintervals, and requires each $\mathcal{Z}(\beta_{k-1})/\mathcal{Z}(\beta_k)\approx\epsilon$, then one has $\epsilon\approx (1/y)^{1/n}$. The corresponding Monte Carlo steps before and after the division scale with $\mathcal{O}(y^2)$ and $\mathcal{O}(n/\epsilon^2)=\mathcal{O}(ny^{2/n})$, respectively. For example, if $y=10^{10}$ as the general order for EE on the lattice model, a crudely calculation needs $\mathcal{O}(10^{20})$ MC steps. If one has only $n=10$ subintervals, the number of the MC steps just decreases to only  $\mathcal{O}(10^3)$. Therefore, the incremental methods could in principle reach exponential magnitude of the increase of algorithm efficiency.

\section{Convergence and optimization for $\Delta U$}
In this section, we aim to investigate the optimization by tuning the control parameter in the new algorithm. As above, one of the most important parameter, which also serves as the essence of an algorithm, is $\Delta U$, expressed in the example of Hubbard model. Considering that $\Delta U$ is always small, it is the reason why we regard the method as the new incremental algorithm in parameter space. If $\Delta U$ is large, the algorithm returns back to the analogue of Grover's original method, which is also in face of the exceptional values problem. However, much dense $U$ values may be waste of resources. If one is only attracted to the behavior of one point in parameter space, for example the behavior near QCP, the possible way to avoid waste, on the premise of the correctness of the algorithm is by setting unfixed $\Delta U$ along the whole track in the parameter space, or calculating EE at certain parameters near the QCP by means of other methods. However, considering the case, when the consecutive behavior of EE in parameter space raises one's interest, one could only make use of the former way to give a proper division of $\Delta U$. To simplify the study, we consider the $\Delta U$ as a constant for whole path, but serves as a tuning parameter to optimize the new algorithm.

\begin{figure}[htp!]
    \begin{minipage}[htbp]{0.4\columnwidth}
        \centering
        \includegraphics[width=\columnwidth]{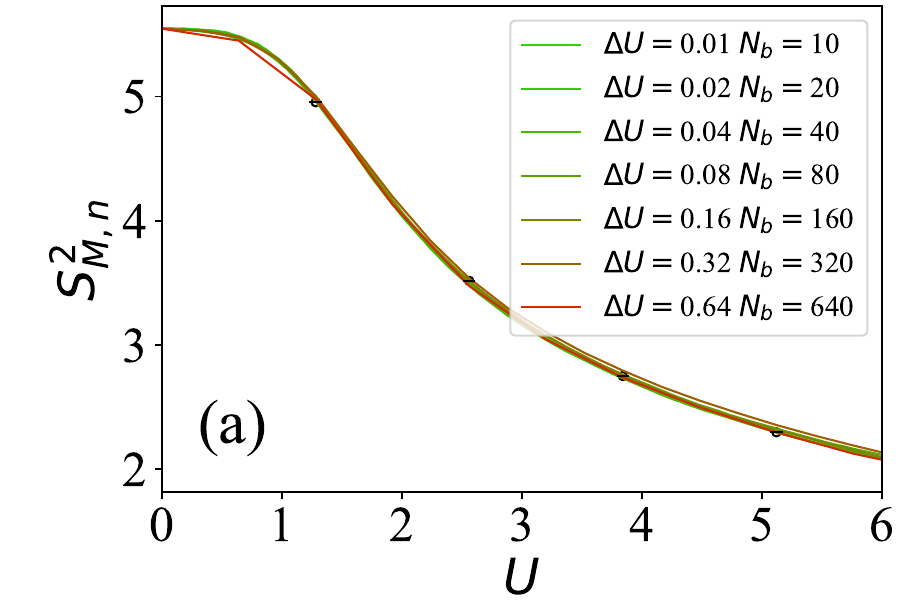}
    \end{minipage}
    \begin{minipage}[htbp]{0.4\columnwidth}
        \centering
        \includegraphics[width=\columnwidth]{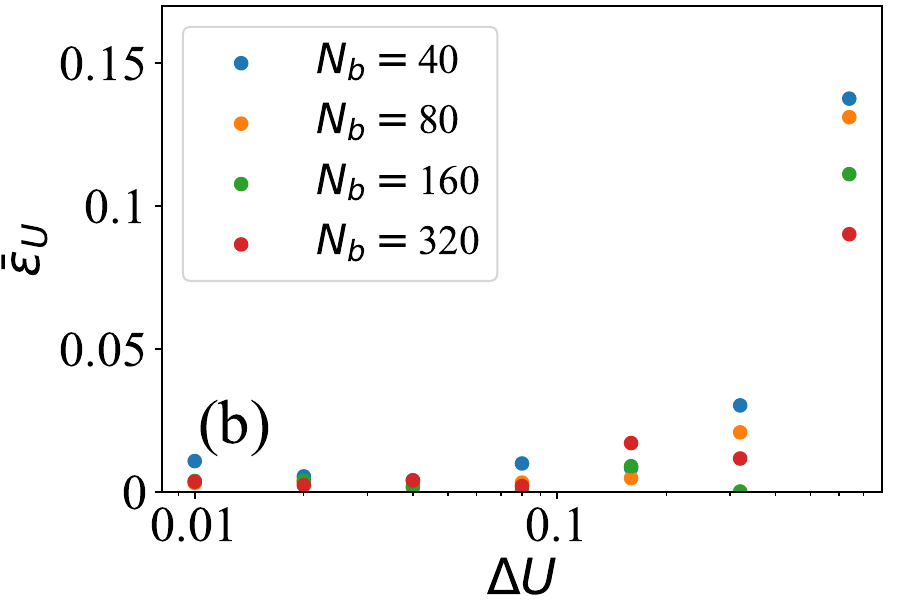}
    \end{minipage}
  \caption{(a) EE results of new algorithm $S^{(2)}_{M,n}$ for various $\Delta U$ at $L=4,\beta=4$. The color change from green to red, corresponding the increasement of $\Delta U$. The black dots are calculated using Grover's method at $U=1.28,2.56,3.84,5.12$. For justice, we also increase $N_b$ to prove the usage of same computation resource. We observe that the red curves has large deviation from the Grover's result. All the simulations are done in presence of TBC. (b) $\epsilon$ versus $\Delta U$ and $N_b$ at $L=4,\beta=4$. As we expect, the deviation decreases as the interval becomes small, or the measurement becomes large. The deviation is smaller than $10^{-3}$ when we choose $\Delta$ smaller than 0.1, and basically remain unchanged if we continue reducing $\Delta U$, indicating our proper choice for $\Delta U$ in the computation of the paper. Note there are 200 sweeps in each bin for the smallest $\Delta U$. The number of the sweep is linearly proportion to $\Delta U$.} 
  \label{fig:opt}
\end{figure}

Firstly, we check for the validity for EE results by varying $\Delta U$.  Taking the calculation from $U = 0$ to $6$ as example, supposed we have the same computing resources, when increasing $\Delta U$, it is fair to enhance the number of samples correspondingly. In Fig.~\ref{fig:opt}(a), we show the results from different $\Delta U$ and compare with the results from Grover\cite{grover2013entanglement}. At small $\Delta U$, we observe a good consistency between the new algorithm and the Grover's method. As $\Delta U$ increase, the EE curves gradually deviate from the data points by Grover's method, raising the challenge for the data correctness. Besides, we add the number of bins, named $N_b$ for each $\Delta U$ calculations to study the convergence to Grover's result. 

To give a quantitative description, we define the deviation, named $\epsilon$, between the results from two methods at certain $U$ points. $\epsilon = | S^{(2)}_{M, n}-  S^{(2)}_{M, g} |/ S^{(2)}_{M, g} $, where $S^{(2)}_{M,n}$ and $S^{(2)}_{M,g}$ represents the EE results from the new algorithm and Grover's method, respectively. We do enough measurement to prove the accuracy for Grover's method, since it serves as the benchmark data. We further define $\bar \epsilon$ to describe the average deviation for many $U$ points.

We plot the value of $\bar \epsilon$, as a function of $\Delta U$ and $N_b$, shown in Fig.~\ref{fig:opt}(b). As we expect, $\bar \epsilon$ gradually converges to 0 as $\Delta U$ decrease. In comparison to $\Delta U$, increasing $N_b$ only has small influence on the deviation. Such quantitative study shows the deviation depends more on $\Delta U$, instead of $N_b$. Above findings inspires us to reduce $\Delta U$ to exploit the advantages for the incremental method. And the value we choose for $\Delta U$ is refer to such analysis, where the deviation is small enough to reach convergence. 

\section{Projection issue}
An intrinsic principle of the projection QMC is acting on $\exp(-\Delta \tau H)$ on the imaginary time ceaselessly to eliminate the weight of excitation state, where the projection length $\beta$ controls the degree of ground state proximity. Such process will become difficult to handle when the gap between the ground state and the first excitation tends to zero, since quite large $\beta$ is needed to reach the exact ground state. It is still acceptable if the gap is algebraically small, However, for Hubbard model at small $U$ limit it is exponentially small which will cause a problem. Indeed, the condition applies to the square lattice Hubbard model near $U=0$, where the gap diverge as $\sim e^{-\beta \sqrt{\frac{t}{U}} }$. Therefore, the calculation at small $U$ region may be {\it unfaithful} compared with real ground state calculation. As a result, the previous study for EE is carried deep in the insulating phase, i.e., large $U$ limit, to get the favorable fitting result\cite{pan2023stable}. Except for $\beta$ and $U$, the trial wavefunction also bears on the how well the projection performance ( See the section of Twisted Boundary Conditions for details ). Therefore, the projection length $\beta$, serving as a tuning parameter, is adjusted to large enough to reach the ground state as close as possible for various choice of trial wavefunction and $U$s. 

\begin{figure}[htp!]
    \centering
    \includegraphics[width=0.4\columnwidth]{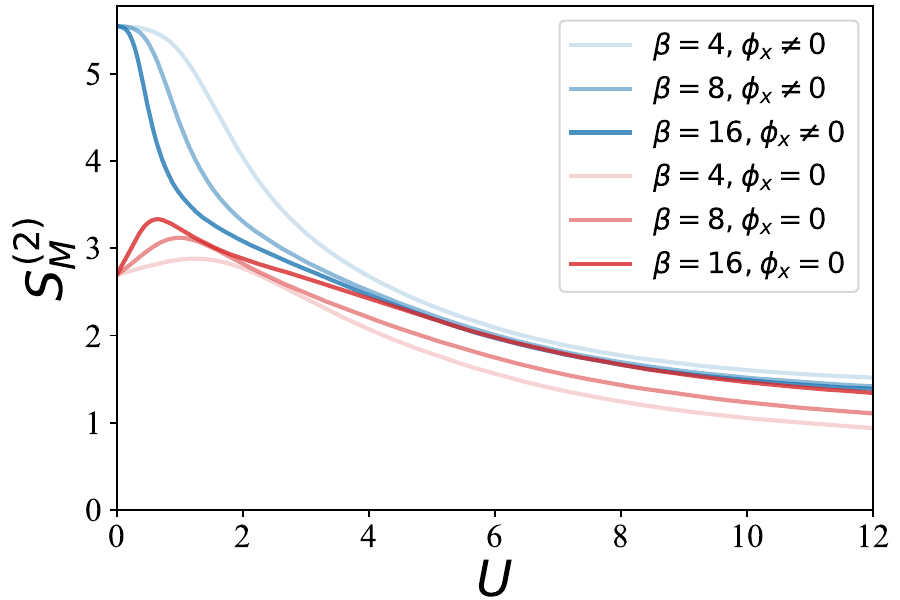}
    \caption{EE results from various projection length and trial wave function at $L=4$. The choice of trial wavefunction is controlled by the TBC, colored by blue (no TBC) and red ($\phi_x = 0.00001$). The gradation of color reflects the projection length. At $U=0$, the distinct TBC leads different EE values. As $U$ becomes large, various curves converge. The two curves with darkest color are most closed to each other, indicating that at large $U$, $\beta=16$ is enough to expose the value of ground state EE.} 
  \label{fig:proj}
\end{figure}

In Fig.~\ref{fig:proj} , we compute EE with various projection lengths and different twisted boundary conditions (TBC). At $U=0$, the EE has only two values depends on the presence of the TBC. In other words, the projection operation does not influence the results for the free fermion. At small $U$, where projection makes a difference, leading to all different EEs. This indicates that the projection length is inadequate, under which condition the wavefuntion after projection differs a lot from the ground state. Nevertheless, we find at large $U$, EE data with same $\beta$ but different trial wavefunctions gradually coincide as $\beta$ increases. The two curves with the darkest color show results of $\beta=16$, which are close to each other, expressing that such projection length is large enough to generate ground state properties regardless of the trial wavefunction. 


\section{Ground state wavefunction at free fermion limit}

To further explore the projection issue, we offer a simple perturbation theory for Hubbard model defined on sqaure lattice. Here, we focus on the small $U$ region and treat $U$ as a perturbation. Taking $L=4$ as an example, on the half-filling condition, the ground state of free fermion limit have degeneracy due to multi-choice for filling at the FS, shown in Fig.~\ref{fig:TBC}. There are total 400 degenerate states, which constitute a subspace for $2^{32}$-dimension of total Hilbert space. Then we numerically solve the eigenvalue and the eigenvector in this subspace at the presence of $U$, and find a non-degenerate ground state. To write down the explicit form of ground state wavefunction, we choose the particle number basis, $| \uparrow_1 \ \uparrow_2 \ \cdots \ \uparrow_6 \ \downarrow_1 \ \downarrow_2 \ \cdots \downarrow_6 \rangle$, where $1,2,\cdots,6$ represents six momentum points on the FS, and $\uparrow,\downarrow$ are spin up and down index. At each momentum points with one spin flavor, the fermion can occupy or not, expressed as 0 or 1. We do the perturbation at small $U$ and obtain the wavefunction, written as,
\begin{equation}
  | \psi_g \rangle = \frac{1}{\sqrt{20}} \sum_{P}  P(\uparrow)  \otimes \bar P(\downarrow) , 
\end{equation}
where $P$ represents state where three of six momentum points to occupy one particle for each, for example $| 100110 \rangle$, and $\bar P$ is opposite configuration, e.g. $| 011001 \rangle$. There are total 20 choice for the combinations, and $| \psi_g \rangle$ is the equal-weight superposition state of 20 basis wavefunctions. We note $| \psi_g \rangle$ also satisfies the exchange invariance for spin up and down. However, we emphasize the form is unable to be written as the trial wavefunction or $P$ matrix in PQMC, since the wavefunction should be the direct product of the electron wavefunctions of two spins. In real simulation, we could only use other forms of trial wavefunction. For example, we use the TBC to choose the certain wavefuntion ( See the section of Twisted Boundary Conditions for details ), and then do the projection operation to reach the ground state. Thus it is always hard to get ground state EE at small $U$ by PQMC if the gap is small.

\section{Twisted boundary conditions}

In this section, we offer the detailed implementation of  In the initialization process, we use TBC to implement various choices of trial wavefunctions for PQMC program. In general, we expect the choice of trial wavefunction to be as close as possible to the ground state wavefunction, in which condition the associated projection length can be small. Generally speaking, to construct the initial wavefunction, we diagonalize the free fermion Hamiltonian in the momentum space, then find several electron wavefunctions with the lowest eigen-energies satisfying the half-filling condition. In fact, on the calculation of lattice model, such ground state wavefunction could not be unity. The problems come from the momentum points on the FS, which lead to the degeneracy of the ground state. Considering the simplest case, $L=4$ square lattice free fermion model with nearest-neighbor hopping, the FS is of skew square shape, with six k-points on it, depicted in Fig.~\ref{fig:TBC}(a). The half-filling condition demands five electronic eigenstates with negative energies and an additional three among six on the FS. Therefore, the amount of choice, i.e. degeneracy, for $L=4$ case is 20 ( 400 for two spin species ), which gets larger along with the system size. In each calculation, we only select one of such degenerate eigenstates to form the $P$ matrix in PQMC. However, we numerically find the choice of the trial wavefunction varies with the compile environment. Importantly, these 20 eigenstates have different EEs, not to mention the linear combination of these orthogonal eigenstates. The former analysis leads to multi-values of EE in the free limit without TBC in different machine, which seems quite awkward, but in fact an existing problem.       

\begin{figure}[htp!]
    \begin{minipage}[htbp]{0.3\columnwidth}
        \centering
        \includegraphics[width=\columnwidth]{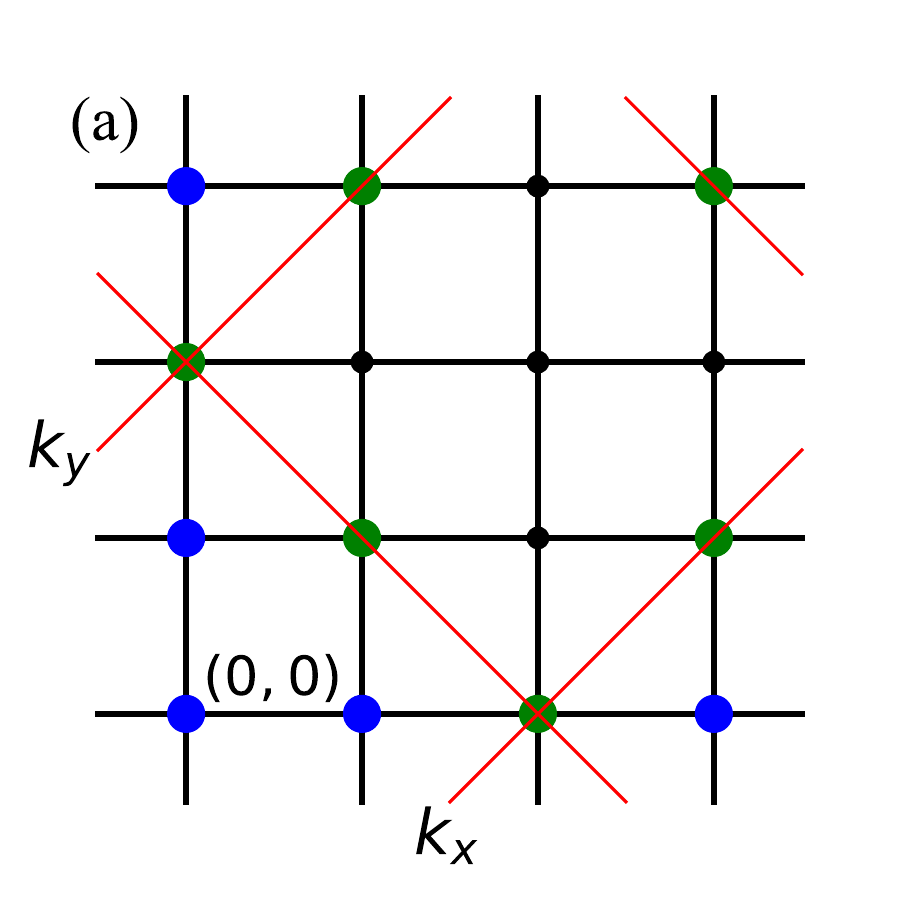}
    \end{minipage}
    \begin{minipage}[htbp]{0.3\columnwidth}
        \centering
        \includegraphics[width=\columnwidth]{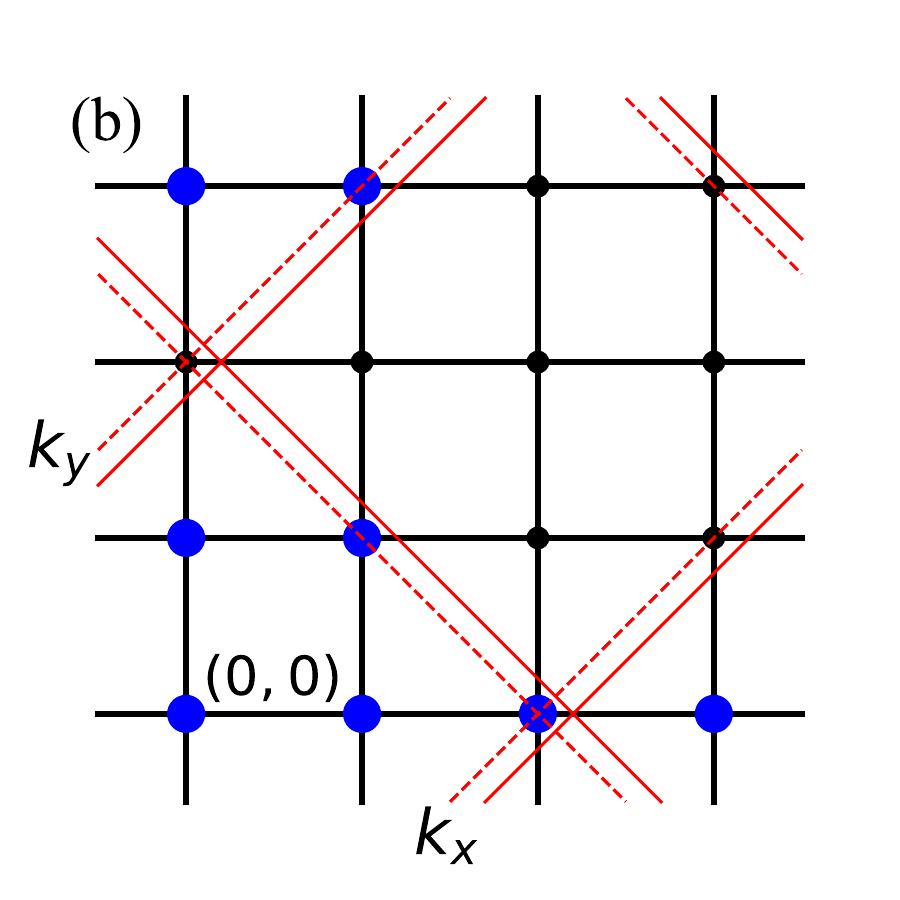}
    \end{minipage}
  \caption{Sketch map of  TBC in the momentum space for $L=4$. (a) In presence of TBC, the FS goes through six momentum points, colored by green. The five blue dots are the momentum points with negative energies. (b) When $\phi_x \neq 0$, the FS translates along the $x$-direction, drawn by the solid line after the translation. Eight momentum points possessing negative energies are colored in blue, which is exactly half of the total number. } 
  \label{fig:TBC}
\end{figure}

An optional way to avoid the multi-choice problem dependent on non-physical factors is to adopt the TBC before diagonalization. The TBC is applied by adding Peries phase factor on each hopping amplitude as Eq.\eqref{eq:S4}. Such condition translates the eigenstates along the direction TBC added in the momentum space. We use $\phi$ to control the translation degree. On the two dimensional square lattice, we add different $\phi$ along $x$ and $y$ direction, and the Hamiltonian with TBC writes, 
\begin{equation}
  H=-t \sum_{i\sigma}   \left( e^{i\phi_x}c_{i \sigma}^{\dagger} c_{i+\hat x \sigma}+ e^{i\phi_y}c_{i \sigma}^{\dagger} c_{i+\hat y \sigma} + \text { H.c. }\right).
  \label{eq:S4}
\end{equation}

As an example, we only add a small $x$-direction twist, i.e., $\phi_y=0, \phi_x = 0.00001$, and show the FS after translation in Fig.~\ref{fig:TBC}(b). There are exact eight negative eigenvalues marked by blue, just half of the total number of momentum points. Therefore, this kind of TBC leads to the half-filling condition, which adapts to all even system size. We calculate EE at $L = 4$ to 16 under such condition, and fit by Eq. (5) in the main text, shown in Fig.~\ref{fig:fitU0}. We numerically find $A$ is close to its thermodynamic limit value 0.5.   

\begin{figure}[htp!]
    \centering
    \includegraphics[width=0.4\columnwidth]{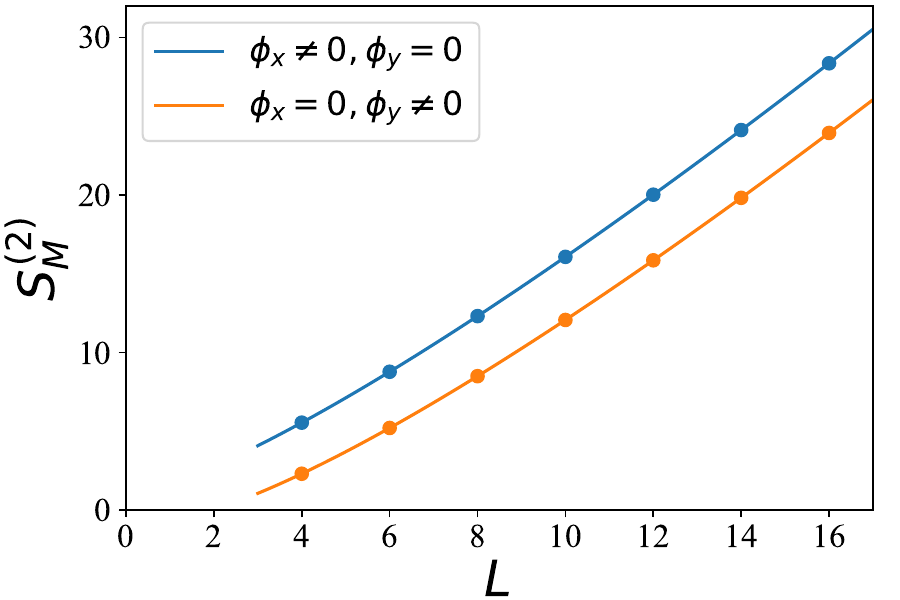}
    \caption{Fitting results of free fermion EE under different TBCs. The blue and yellow line represents the TBC along $x$ and $y$ direction, respectively. We fit two data sets with Eq. (5) in the main text and focus on the leading term coefficient $A$. Here, $A_x = 0.48(3), A_y = 0.51(3)$ for  the TBC along $x$ and $y$ direction. Both values are close to the  thermodynamic limit value 0.5. For $\phi \neq 0$ condition, we give $\phi$ a small but non-zero value, e.g. 0.00001 in the program.}  
  \label{fig:fitU0}
\end{figure}

One could notice that since the entangled region $M$ is unequal for $x$ and $y$ direction, applying $y$-direction twist $\phi_x=0, \phi_y = 0.00001$ of course results in different EE values. Fortunately, we also obtain similar fitting results for $L \ln L$ term coefficient in Fig.~\ref{fig:fitU0}. We conclude that, if the TBC is fixed for all system sizes, or more generally the choice principle of half-filled electron eigenvectors in momentum space, the leading term coefficient of EE could emerge close to the thermodynamic limit. Therefore, even though the ground state EE is not unique in PQMC, we are able to identify the scaling behavior for further analysis. 

Equipped with above techniques, we could get a deep understand for the fitting result in Fig. 3 of the main text.
In the case with TBC ($\phi_x \neq 0$) applied, EE fits well with Eq. (4) in the main text at $U=0$ (blue lines in Fig.~\ref{fig:fitU0}), providing the $A = 0.5$ as expectation in (c) ( See next section for the derivation ). However, closed to the free fermion limit, we observe a plateau in (a) in the main text, which is supposed to be a performance sharing the similarity to its trial wavefunction at $U=0$. Meanwhile, it is found that the goodness-of-fit becomes worse at $U \sim [1,4]$ for both two scaling function forms shown in (d). The reason indeed comes from the inadequate projection at the small $U$ region, shaded by white in (b), where the gap between the first excitation state and the ground state is relatively small. Above problem disappears at large $U$ since the gap becomes larger and the projection is adequate to reach the true ground state, where the fitting results in (e) accords with the area law with Goldstone modes. Nonetheless, the above behaviors of EE originates from the projection methods itself, rather than the discrepancy of our new developed algorithm. 

\section{Widom-Sobolev equation for free fermion limit}

\begin{figure}[htp!]
  \includegraphics[width=0.5\columnwidth]{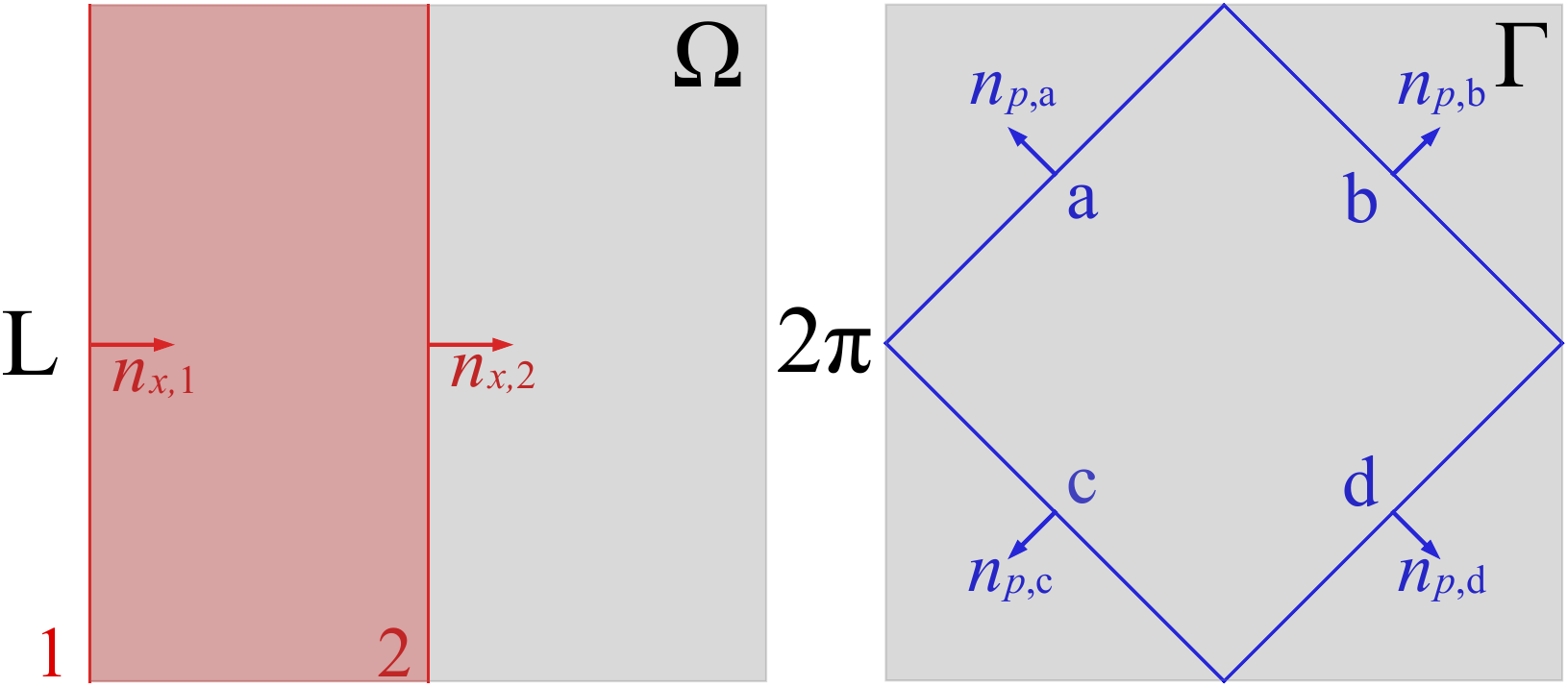}
  \caption{ The sketch map of boundaries of real space subregion in the left panel, and FS in the momentum space. The real space boundary is divided into two parts, labeled $1$ and $2$, where $n_{x,1}$ and $n_{x,2}$ are the associated normal vectors. The FS boundary is divided into four parts, labeled $a,b,c,d$, where $n_{x,a},n_{x,b},n_{x,c},n_{x,d}$ are the associated normal vectors.} 
  \label{fig:WS}
\end{figure}

The scaling behavior of ground state EE in free fermion system have experienced a long study, where the pioneer work was concluded as Widom conjecture\cite{Gioev2006Entanglement}. The crucial discovery is that in presence of the FS, the leading term of EE scales as $L^{d-1} \ln L$, where $d$ is the dimension exceeding the general area law behavior. Latter, Brian proposed a phenomenological analysis for the emergence of $L \ln L$ term\cite{Swingle2010Entanglement}. In brief, for two dimensional system, each point on the FS owns a chiral model contributing to the $\ln L$ term, as described by one dimensional conformal field theory. Since the mode density scale with $L$, EE with FS scales as $L \ln L$. Besides, the leading term coefficient also depends on the shape of FS and subregion. In 2014, Leschke and et al. gave an rigorous proof for the more general version of Widom conjecture, and extended it from smooth functions to a certain class of non-smooth functions, known as the Widom-Sobolev equation\cite{Leschke2014Scaling}. The $n$-order Renyi entropy has 
\begin{equation}
	S^{(n)}_M \sim  \frac{n+1}{24n}  \frac{L^{d-1}\ln L}{(2\pi)^{d-1}} \int_{\partial \Omega} \int_{\partial\Gamma} |\mathbf{n}_x \cdot \mathbf{n}_p| \mathbf{d}S_x \mathbf{d}S_p,
\label{eq:widom}
\end{equation}
where $\partial \Omega, \partial\Gamma$ represents the integral along the boundary of the subregion $M$ and FS. $\mathbf{n}_x,\mathbf{n}_p$  is the unit normal vector with respect to the subregion and FS in the momentum space. $\mathbf{d}S_x$ integrates in the real space with unit length, $\mathbf{d}S_p$ in the momentum space. Note the subregion is chosen as a rectangle, shown in Fig. 1(b) in the main text, where the boundaries only exist along verticle direction due to the period boundary condition. Since the boundary for subregion and FS are all straight, the term $|\mathbf{n}_x \cdot \mathbf{n}_p|$ can be regarded as the projection for two boundaries from $\mathbf{d}S_x$ and $\mathbf{d}S_p$. Therefore, the total integral is divided into the single integral of each boundary, 
\begin{equation}
    \begin{aligned}
    S^{(n)}_M &\sim  \frac{n+1}{24n}  \frac{L^{d-1}\ln L}{(2\pi)^{d-1}} \int_{\partial \Omega} \int_{\partial\Gamma} \mathbf{d}S_x \mathbf{d}S_p \cos(\beta_{x,p}), \\
    & \sim \frac{n+1}{24n}  \frac{L^{d-1}\ln L}{(2\pi)^{d-1}} \int_{1,2} \int_{a,b,c,d} \mathbf{d}S_x \mathbf{d}S_p \cos(\beta_{x,p})
    \label{eq:widom2}
    \end{aligned}
\end{equation}
$\beta_{x,p}$ represents the angle between two boundaries $1,2$, $a,b,c,d$ are boundaries of subregion and FS, respectively. The result of the integral is $8\pi$. Note we have two spin species in the free fermion limit, the final coefficient of $A$ in Eq. (5) is 0.5 for theoretical result.

\end{widetext}

\end{document}